\documentclass[12pt]{article} 
\usepackage{graphicx}
\newcommand{\be}{\begin{equation}}
\newcommand{\ee}{\end{equation}}
\newcommand{\bear}{\begin{eqnarray}}
\newcommand{\eear}{\end{eqnarray}}
\newcommand{\ba}{\begin{array}}
\newcommand{\ea}{\end{array}}

\begin{document}

\begin{titlepage}
\vfill
\begin{flushright}
{\normalsize IC/2009/066}\\
{\normalsize arXiv:0908.4189[hep-th]}\\
\end{flushright}

\vfill
\begin{center}
{\Large\bf Holographic Chiral Magnetic Conductivity }

\vskip 0.3in

{ Ho-Ung
Yee\footnote{\tt hyee@ictp.it}}

\vskip 0.15in

 {\it ICTP, High Energy, Cosmology and Astroparticle Physics,} \\
{\it Strada Costiera 11, 34014, Trieste, Italy}
\\[0.3in]

{\normalsize  2009}

\end{center}

\vfill

\begin{abstract}

We present holographic computations of the time-dependent chiral magnetic conductivity in the framework of gauge/gravity correspondence.
Chiral magnetic effect is a phenomenon where an electromagnetic current parallel to an applied magnetic field
is induced in the presence of a finite axial chemical potential.
Motivated by a recent weak-coupling perturbative QCD calculation, our aim is
to provide a couple of complementary computations for strongly coupled regime which might be relevant
for strongly coupled RHIC plasma.
We take two prototypical holographic set-ups for computing chiral magnetic conductivity; the first model    is Einstein gravity with $U(1)_L\times U(1)_R$ Maxwell theory, and our second set-up is based on the Sakai-Sugimoto model in a deconfined and chiral symmetry restored phase. While the former takes into account full back-reaction while the latter not, the common feature is an important role played by
the appropriate 5-dimensional Chern-Simons term corresponding to the 4-dimensional axial anomaly.

\end{abstract}

\vfill

\end{titlepage}
\setcounter{footnote}{0}

\baselineskip 18pt \pagebreak
\renewcommand{\thepage}{\arabic{page}}
\pagebreak

\section{Introduction}

For certain strongly coupled dynamics of gauge theories,
gauge/gravity correspondence has become a useful method to study the problems, alternative
to the conventional techniques such as perturbation theory.
Although it was originally developed in the case of large color $N_c$ and strong t'Hooft
coupling $\lambda=g_{YM}^2 N_c$ limit, it has given us a considerable amount of new insights
on generic strongly coupled gauge theories, and in many occasions its predictions
capture, at least qualitatively, right physics for otherwise difficult non-perturbative phenomena.
Of particular interests are of course possible applications to QCD. As QCD coupling runs to
a large value at low energy, it is a logical hope that some low energy QCD phenomena
which are hard to be explained by other means may have explanations in holographic QCD.
It is at least worthwhile to study the problems in the framework to see what it pre/post-dicts
and also to compare with other known methods.

Not only low energy QCD but also finite temperature deconfined quark-gluon plasma has
been an active area of applications of gauge/gravity correspondence. The main motivation
is the experimental finding at RHIC indicating a strongly coupled phase of quark-gluon plasma.
Moreover, one may expect more sensible connections between finite temperature phases of different
gauge theories and the QCD plasma, because at finite temperature fermions and scalar bosons get
effective masses and become less relevant than the universal gauge field dynamics.
Hydrodynamics would be a right place for searching for some universality as it is describing precisely the
long wavelength transport dynamics for which these massive modes may decouple.
There has been an enormous amount of recent works studying hydrodynamics and transport coefficients
in the gauge/gravity correspondence ( see refs.\cite{Son:2007vk,Rangamani:2009xk} for reviews),  although it still seems to remain as a fruitful area of further research.

In this work, we will study one more example to the plethora of holographic QCD applications :
the computations  of chiral magnetic conductivity \cite{Fukushima:2008xe,Kharzeev:2009pj} at finite frequency.
This is motivated by a recent work of Kharzeev and Warringa \cite{Kharzeev:2009pj} which computes the time-dependent chiral
magnetic conductivity in 1-loop perturbative QCD, aiming at a weakly coupled phase of QCD plasma at very high temperature.
Chiral magnetic effect is the phenomenon where an electromagnetic current is induced parallel to the applied magnetic field
in the presence of non-zero chiral density (i.e. unbalance between positive and negative helicity particles), and it is one kind of chiral-anomaly originated effects. Chiral magnetic conductivity
is the proportionality coefficient of the induced current to the magnetic field.
In a QCD plasma such as the RHIC experiment, finite local chiral density can be generated by sphaleron fluctuations, and a large
magnetic field may also appear in off-center collisions along the direction of angular momentum.
Consequently, chiral magnetic effects may play some role in the subsequent plasma dynamics, and
this was the motivation of the above authors.
However, as RHIC plasma seems to become rather strongly coupled shortly after the collision, one needs other methods to complement the previous weak coupling calculation. Gauge/gravity correspondence would be a worthwhile try for this purpose. Lattice simulations for chiral magnetic effects for {\it static} but arbitrary magnitudes of magnetic field are given in refs.\cite{Buividovich:2009ih}.

In fact, the possibility of using holography for computing chiral magnetic conductivity was
first pointed out by Rebhan, Schmitt and Stricker in ref.\cite{Rebhan:2008ur}. Moreover, Son and Surowka
recently computed a similar quantity (they call $\xi_B$) in the {\it static} magnetic field case \cite{Son:2009tf}.
See also the work by Lifschytz and Lippert \cite{Lifschytz:2009si} for other interesting phenomena related to chiral anomaly in the holographic set-up.
However, it seems that there has been no study on the frequency dependent behavior of the chiral magnetic conductivity, which would be relevant in hydrodynamic simulations of the RHIC experiment.
This will be our main focus and results in this paper.

As the precise holographic model of large $N_c$ QCD is not known up to now,
our objective is
to set-up a couple of consistent holographic frameworks to compute time-dependent chiral magnetic conductivity, and to provide the results based on those.
Our first holographic model is the 5-dimensional Einstein gravity with a negative cosmological constant
coupled to two $U(1)$ Maxwell fields; $U(1)_L\times U(1)_R$. As should be clear from the notation, these
two 5-dimensional gauge theories correspond to the 4-dimensional
$U(1)_L\times U(1)_R$ chiral symmetry that we are interested in.
We will consider an {\it exact} Reisner-Nordstrom black-hole solution charged only under the axial $U(1)_A=U(1)_L-U(1)_R$ to represent a finite temperature phase with non-zero chiral density\footnote{See refs.\cite{Son:2006em} for previous studies of using this solution.}.
The "electromagnetism", of which we will turn on an external magnetic field and also read off the induced current, is the vector part $U(1)_{EM}=U(1)_L+U(1)_R$.
The second holographic set-up that we will study is based on the Sakai-Sugimoto model \cite{Sakai:2004cn}, which
seems to be closer to the realistic QCD in quenched approximation.
We will consider its deconfined, chiral symmetry restored phase with one flavor $N_F=1$, whose chiral symmetry is also
$U(1)_L\times U(1)_R$. The effective 5-dimensional $U(1)_L\times U(1)_R$ gauge theory
is a Dirac-Born-Infeld action and the back-reaction of it to the background geometry is consistently
neglected in quenched/probe approximation. This should be contrasted to the first holographic model where
the full back-reaction is taken into account.

In both set-ups, the common feature is the 5-dimensional Chern-Simons couplings for $U(1)_L\times U(1)_R$
gauge theory living in the holographic 5-dimensional bulk, corresponding to the 4-dimensional chiral anomaly. As this is more or less dictated uniquely by the anomaly structure, it is a universal
feature, and the results directly related to it should be taken as robust.
The only difference between the two models is the details of the background metric and how to treat
the back-reaction of the finite axial/chiral density. One can take our set-ups as two
prototypical examples whether or not we consider the back-reaction of the axial/chiral density.

{\it Note added : }
Shortly after this paper, there appeared an interesting observation in ref.\cite{Rebhan:2009vc} regarding the correct identification of holographic currents in the presence of 5D Chern-Simons terms, which is relevant in our computation of chiral magnetic conductivity. Ref.\cite{Rebhan:2009vc} computed zero frequency chiral magnetic conductivity in the Sakai-Sugimoto model taking into account additional contributions coming from
these modifications. We will briefly summarize these modifications here, that will correct our currents
we used in the text {\it by a constant, frequency-independent shift}.
At the end, we will also mention a few puzzles that still remain even if we take this modification into account, which should be resolved in the near future.

Firstly, as ref.\cite{Rebhan:2009vc} observed, one easily derives that the variation of 5D Chern-Simons term in the 5D action gives us additional contribution to the currents,
\be
\Delta_{CS}J^\mu_L=-{N^{eff}_F N_c\over 24\pi^2}\epsilon^{\mu\nu\rho\sigma} (A_L)_\nu (F_L)_{\rho\sigma}\quad,\quad
\Delta_{CS}J^\mu_R=+{N^{eff}_F N_c\over 24\pi^2}\epsilon^{\mu\nu\rho\sigma} (A_R)_\nu (F_R)_{\rho\sigma}\quad,
\ee
which should be added to our currents in the text according to AdS/CFT dictionary. Note that the gauge fields appearing on the right-hand sides are {\it external} UV boundary fields, without any component of
subleading dynamical piece. This is one characteristically different property of these additional contributions, which are of local, contact-term type,
compared to the one from the subleading piece in the text. The total currents would be then the sum of the two if we accept these modifications.
The above gives to the EM current $j_{EM}=e(j_L+j_R)$ an extra piece
\be
\Delta_{CS} j_{EM}={e^2 N^{eff}_F N_c\over 12\pi^2}\epsilon^{\mu\nu\rho\sigma}\left((A_a)_\nu (F_{EM})_{\rho\sigma} +(A_{EM})_\nu (F_{a})_{\rho\sigma}\right)\quad,
\ee
where the external EM and axial gauge potentials are defined by $A_L=eA_{EM}-A_a$ and $A_R=eA_{EM}+A_a$.
In our situation, we turn on the time component of axial potential $(A_a)_0=\mu_a$ as a chemical potential for the axial charge, and also an external EM magnetic field $B^3_{EM}=(F_{EM})_{12}$ along $x^3$ direction, to define chiral magnetic conductivity. From the above, the modification of chiral magnetic conductivity would therefore be
\be
\Delta_{CS} \sigma = -{e^2 \mu_a \over 6\pi^2} \left(N_F^{eff} N_c\right)\quad.
\ee
Note that due to the local nature of the above modifications, the shift is simply a constant without any momentum or frequency dependence of the probe field $B_{EM}$.
Therefore, our plots can simply be shifted by this constant amount without a need for re-computations.

This is one story, while ref.\cite{Rebhan:2009vc} went further to propose an interesting observation.
They realized that even after the above modification, the EM current is not strictly conserved in the presence of external axial potential $A_a$, that is, one can find from the 5D equations of motion that
\be
\partial_\mu J_{EM}^\mu = -{N_F^{eff} N_c\over 24\pi^2} \epsilon^{\mu\nu\rho\sigma}(F_{EM})_{\mu\nu} (F_a)_{\rho\sigma}\quad.
\ee
The details can be found in ref.\cite{Rebhan:2009vc}, but we only mention that they added an additional {\it local counter-term}, called Bardeen-term, in the regularized holographic effective action, to remedy
this non-conservation. This additional counter-term, which can be added on the UV boundary as a different holographic renormalization prescription, gives us additional contribution to the current of similar type
as above from the 5D Chern-Simons term. As can be easily expected, this contribution is also of local type
and its contribution to the chiral magnetic conductivity is again a simple constant without any frequency dependence.

If we choose to include this too, in total the induced EM vector current $j_{EM}=e(j_L+j_R)$
receives additional contribution
\be
\Delta j_{EM}^\mu = {e^2 N^{eff}_F N_c\over 4\pi^2}\epsilon^{\mu\nu\rho\sigma}(A_A)_\nu(F_{EM})_{\rho\sigma}\quad,
\ee
which gives us a {\it constant} shift in the chiral magnetic conductivity by
\be
\Delta \sigma = -{e^2 \mu_a \over 2\pi^2} \left(N_F^{eff} N_c\right)\quad,
\ee
that is precisely minus of the zero frequency value $\sigma(0)$ we obtain in this paper, so that
the zero frequency chiral magnetic conductivity in their prescription vanishes.
As they pointed out, there is no a priori reason for vanishing chiral magnetic conductivity when the vector current is strictly conserved. In fact, because the additional contribution is a frequency-independent shift, the real part of our resulting Figures \ref{first},\ref{first2},\ref{second}, and \ref{second2} are simply shifted down by $\Delta \sigma$, and there is non-zero chiral magnetic conductivity at finite frequency even in their prescription. As the magnetic fields relevant in RHIC experiments are time-dependent, chiral magnetic conductivity will still be at work, but with a different detailed prediction.

Finally, let us mention a few seemingly puzzling aspects of the currents even after we take into account Chern-Simons contributions.
First, combined with our resulting plots, the chiral magnetic conductivity goes to a constant value when $\omega\to\infty$. Dynamically this doesn't make sense, because the medium cannot respond to the perturbation which is arbitrary fast. If we use only subleading piece as we did in the text, we do get
a nice damping when $\omega\to\infty$. This seems to indicate there might be something we are still missing at the moment. Another point, which is probably related to the first point, is that
when we take a variation of 5D Chern-Simons action, one also gets a contribution from the IR boundary, which in our case is the horizon. It could be that one has to consider also the IR boundary contributions
to resolve the first puzzle, which we hope to clarify in the near future.

In summary, there indeed is an issue regarding what is the correct holographic current in the presence of 5D Chern-Simons term as we briefly reviewed current proposals.
At the moment, arguably there seems no definite answer for that. However, the purpose in the present paper
is to study frequency dependence of chiral magnetic conductivity, and because the present differences between different proposals are all {\it constant shifts}, our main results can easily accommodate the
future resolving the issue.

\section{ A quick review on physics of chiral magnetic effect}

A 4-dimensional field theory at finite temperature that gives rise to the chiral magnetic effect, such as chiral symmetry restored phase of QCD plasma, has the following basic ingredients

\begin{itemize}
\item
There are two chiral $U(1)$ symmetries $U(1)_L$ and $U(1)_R$, each having a non-zero triangle anomaly
with the same magnitude but with opposite sign. Equivalently, if one weakly gauges these symmetries
by coupling to non-dynamical gauge fields $A_L$ and $A_R$ respectively, their current conservation laws
are violated by
\bear
\partial_\mu j^\mu_L &=& {N^{eff}_F N_c\over 32 \pi^2} \epsilon^{\mu\nu\alpha\beta}(F_L)_{\mu\nu} (F_L)_{\alpha\beta}\quad,\nonumber\\
\partial_\mu j^\mu_R &=& -{N^{eff}_F N_c\over 32 \pi^2} \epsilon^{\mu\nu\alpha\beta}(F_R)_{\mu\nu} (F_R)_{\alpha\beta}\quad,\label{anomaly}
\eear
where $N^{eff}_F$ is the effective number of flavors counted as fundamental representations of the color $SU(N_c)$, and $F_{L,R}$ are field strengths of $A_{L,R}$.

\item
Turn on a finite chemical potential for the axial $U(1)_A$ whose current is
\be
j_A=-j_L+j_R\quad,
\ee
while keeping the system neutral under the vector "electromagnetic" $U(1)_{EM}$
\be
j_{EM}=e\left( j_L +j_R \right)\quad,
\ee
where $e$ is the electromagnetic coupling constant.
In the language of weakly-gauging symmetries, these correspond to
\be
A_L= e A_{EM} -A_A\quad,\quad
A_R= e A_{EM} +A_A \quad.\label{trans}
\ee
Note that $U(1)_{EM}$ is anomaly-free
and it is consistent to let $A_{EM}$ be a real dynamical gauge theory as the Nature does.

\item
Apply a homogeneous but possibly time-dependent magnetic field of $A_{EM}$ along say $x^3$ direction,
\be
B_{EM}^3 = (F_{EM})_{12} = B(\omega) e^{-i\omega t}\quad,
\ee
which should be treated as an external perturbation to the system. Then the chiral magnetic effect
induces the electromagnetic current $j_{EM}^3$ parallel to $B_{EM}^3$
\be
j_{EM}^3 = j(\omega) e^{-i\omega t} \equiv \sigma(\omega) B(\omega) e^{-i\omega t}\quad.
\ee
The $\sigma(\omega)$ is the chiral magnetic conductivity \cite{Fukushima:2008xe,Kharzeev:2009pj}.

\end{itemize}

An intuitive explanation on the microscopic origin of this phenomenon was given in ref.\cite{Kharzeev:2009pj}.
For simplicity, consider a free massless one flavor of quarks $(q_L,q_R)$ with unit electromagnetic charge.
We emphasize that masslessness is important to have chiral symmetry.
Upon quantizing $q_L$, one gets a particle of negative helicity (meaning that its spin is opposite
to its momentum) as well as its anti-particle of positive helicity with a negative electromagnetic charge, and
{\it vice versa} for $q_R$. Due to the Wigner-Eckart theorem, the magnetic moment of an elementary particle
should be proportional to its spin, and for positively charged particles/anti-particles the magnetic moment
is in the same direction of the spin, while for negatively charged ones it is opposite to the spin.
Now imagine applying an external electromagnetic magnetic field, then the magnetic moments of particles/anti-particles will tend to align along the direction of the applied magnetic field.
Because the magnetic moment, the spin, and the momentum are correlated with each other as described above,
one can easily deduce the following pattern of responses
\begin{itemize}
\item
Positively charged particles as well as negatively charged anti-particles from $q_L$ tend to move in reverse direction to the magnetic field. Let's denote them as $(q^+_{-1/2}, q^-_{+1/2})$ where
the upper index represents the charge and the lower the helicity.

\item
Positively charged particles as well as negatively charged anti-particles from $q_R$ tend to move
towards the same direction as the magnetic field. We denote them as $(q^+_{+1/2}, q^-_{-1/2})$.
\end{itemize}
Then, having a finite chemical potential for the axial $U(1)_A$ symmetry which might be achieved by
local sphalerons means that the number of positive helicity states is larger than the number of negative
helicity states
\be
N\left(q^-_{+1/2}\right)+N\left(q^+_{+1/2}\right) > N\left(q^+_{-1/2}\right)+N\left(q^-_{-1/2}\right)\quad.
\ee
In conjunction with the above discussion, observe that the left-hand side of the above inequality induces a positive electromagnetic current
along the magnetic field, while the right-hand side would contribute to a current in opposite direction to the magnetic field, so that the above inequality tells us there would be a net positive electromagnetic
current induced along the magnetic field : this is the chiral magnetic effect.

Although it is not absolutely necessary, the electromagnetic neutrality that we require for simplicity implies that
\be
N\left(q^+_{-1/2}\right)-N\left(q^-_{+1/2}\right)= -\left(N\left(q^+_{+1/2}\right)- N\left(q^-_{-1/2}\right)\right)\quad,
\ee
where the left-hand side is proportional to the chemical potential for $q_L$ (or $U(1)_L$) and the
right-hand side is the minus of the chemical potential for $q_R$ (or $U(1)_R$), which implies $\mu_R=-\mu_L=\mu_A$ with $\mu_A$ being the axial chemical potential we turn on in the background.
It might be an interesting future direction to generalize our computations to the cases with non-zero electromagnetic charge density too.

\section{Retarded response (Green's) function in Eddington-Finkelstein coordinate}

Before presenting holographic calculations of chiral magnetic conductivity, let us make a short
digression to explain our method of obtaining retarded response function in a black-hole background.
There is by now a well-established procedure to compute any retarded response function in gauge/gravity correspondence {\it in linear response approximation} \cite{Son:2002sd}, and our results will also belong to this category.
As we are interested in the finite frequency $\omega$, the method based on derivative expansions such as
those of ref.\cite{Bhattacharyya:2008jc}, although it is fully {\it non-linear}, is not suitable for our purpose, as one necessarily truncates the expansion at some finite order of derivatives and the results lose its validity at high $\omega$.
On the other hand, the linear response approach can be used at any frequency $\omega$ while it
requires the driving source to be small for the linear approximation to be valid.
It seems both approaches have pros and cons.

In linear response approach, one perturbs the system by an external source $B$ coupled to an operator $J$ of the theory, and the first-order perturbation theory tells us that the response in the expectation value of the operator $\langle J \rangle$ is given by the convolution of the source with the retarded Green's function
\be
\langle J(x) \rangle= - \int d^4 x' \, G_R(x,x') B(x')\quad.
\ee
Typical interests, including our present work, are therefore the computations of the retarded response function $G_R$. In the language of quantum mechanics, the retarded response function $G_R$ has an expression
\be
G_R(x,x')= (-i)\theta(t-t'){\rm tr}\left( e^{-\beta H}[J(x), J(x')]\right) \quad, \label{GR}
\ee
and it is one kind of 2-point Green's function in Minkowski signature. Because gauge/gravity correspondence is typically formulated as a prescription to compute precisely the Green's functions or correlation functions, one might hope to compute $G_R$ rather easily by simply applying the suitable gauge/gravity dictionary. However, this "suitable" dictionary turns out to be rather non-trivial in Minkowski signature,
especially retarded Green's function in the presence of black-hole horizon, issues being whether or not one should treat the horizon as boundary, etc. The issues have been settled by now, and one has a definite
well-defined way of computing $G_R$ holographically in linear response theory \cite{Son:2002sd}.

However, looking back the original motivation of studying the causal response in the presence of driving
external source, the quantum mechanical expression (\ref{GR}) which is a language of 4-dimensional field theory side in fact seems unnecessary. What one is interested is simply the resulting $\langle J(x) \rangle$
causally responding to the external source $B(x)$ in the Minkowski evolution of the 5-dimensional holographic dual theory. The fact that $G_R$ has a field theory interpretation of 2-point correlation function is not needed to find the answer, {\it if one can directly solve the Minkowski dynamics in 5-dimensions in the presence of external source $B(x)$, because the expectation value $\langle J(x) \rangle$
is also encoded in the resulting 5-dimensional solution as a normalizable mode in the near boundary
expansion according to the holographic renormalization} \cite{de Haro:2000xn}.
In other words, once we accept the results of standard holographic renormalization {\it a la} ref.\cite{de Haro:2000xn},
which states that the expectation values can simply be read off from the near boundary expansion
of the 5-dimensional fields {\it even in the presence of black-hole horizon and in Minkowski signature}, we can by-pass the issue
of calculating retarded Green's function, by simply solving the Minkowski equation of motions
with "physically obvious" in-coming boundary condition at the black-hole horizon. With the UV boundary
condition also fixed by the given source $B(x)$, this uniquely determines the 5-dimensional solution and from it one can
directly read off $\langle J(x) \rangle$.
The important question is whether this way gives us the same results consistently to those obtained by the well-established
method of computing retarded Green's functions.
For a massless scalar, it is checked to be true, and we conjecture it is always true.

\begin{figure}[t]
\begin{center}
\includegraphics[width=12cm,height=7cm]{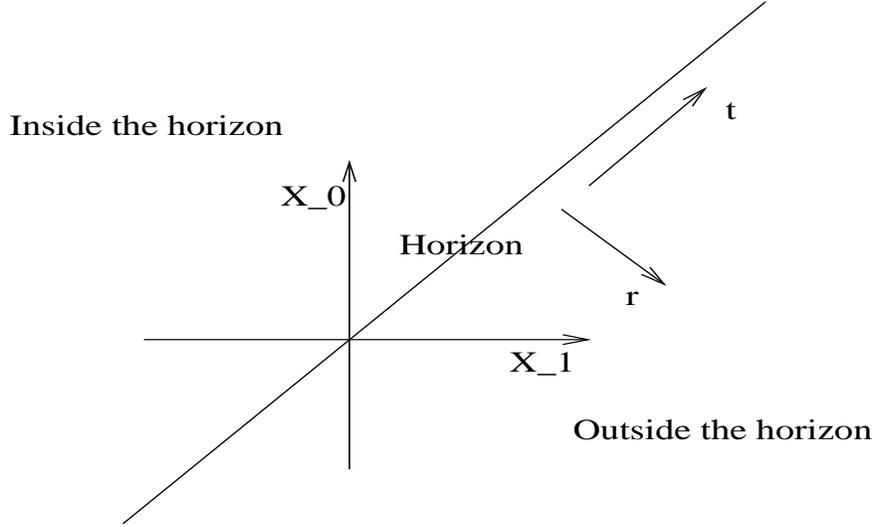}
\caption{\small A schematic description of the horizon in the Eddington-Finkelstein coordinate. }
\label{ed}
\end{center}
\end{figure}

Turning to a practical side, we will adopt the Eddington-Finkelstein coordinate system for a given background black-hole geometry in which the metric takes a form
\be
ds^2= -r^2 V(r) dt^2 +2 dr dt + r^2 \sum_{i=1}^3 \left(dx^i\right)^2\quad,
\ee
where the horizon is located at the position $V(r_H)=0$. The motivation is that the in-coming boundary condition at the horizon can be easily fulfilled in this coordinate by simply imposing only
{\it regularity} at the horizon on the solutions.
To see this more clearly, consider a mode with frequency $\omega$ along the time-like Killing vector $\partial\over\partial t$
\be
\phi(r,t) \sim e^{-i\omega t} f(r)\quad.\label{mode}
\ee
Near the horizon where $V(r)\approx 0$, the metric becomes
\be
ds^2 \sim 2 dr dt +ds^2_{T} \quad,
\ee
where $ds^2_{T}$ is along the transverse directions $x^i$ that is not important in the discussion.
Therefore, the $(r,t)$ coordinates become two local null directions near the horizon,
and in terms of the more standard local flat coordinates, say $(X^0,X^1)$, they are written as
\be
t={1\over \sqrt{2}}\left(X^0 +X^1\right)\quad,\quad r={1\over \sqrt{2}}\left(-X^0 +X^1\right)\quad,
\ee
so that $ds^2 \sim -\left(dX^0\right)^2 + \left(dX^1\right)^2 +ds^2_{T}$. In writing the above, we re-parameterize $r$
such that $r=0$ is the location of horizon.
See the Figure \ref{ed} for a schematic definition of $(X^0,X^1)$, and the region of $X^0 > X^1$
is inside the horizon while $X^0<X^1$ corresponds to the region outside the horizon.
Hence, the mode (\ref{mode}) on which we impose only {\it regularity} at the horizon $r=r_H$
has a near horizon behavior
\be
\phi \sim e^{-i\omega t} f(r_H) = e^{-i{\omega\over\sqrt{2}}\left(X^0+X^1\right)} f(r_H)\quad,
\ee
which is {\it automatically} in-coming towards the region inside the horizon. Note that the conclusion
doesn't depend on the signature of $\omega$.

Another way of understanding this is to go back to the more conventional form of the black-hole metric
\be
ds^2 = -r^2 V(r) dt_*^2 + {dr^2\over r^2 V(r)} +ds^2_{T} \quad,
\ee
achieved by a relation
\be
t_*= t - \int_\infty ^r \,{dr' \over (r')^2 V(r')}\quad.
\ee
Note that the time-like Killing vector retains the same expression
\be
{\partial \over \partial t}= {\partial \over \partial t_*}\quad,
\ee
and a generic mode with frequency $\omega$ would be
\be
\phi(r,t_*) = e^{-i\omega t_*}f_*(r) = e^{-i\omega t}\left(e^{i\omega \int_\infty ^r \,{dr' \over (r')^2 V(r')}}f_*(r)\right) \equiv e^{-i\omega t} f(r)\quad,
\ee
so that by definition
\be
f_*(r)=e^{-i\omega \int_\infty ^r \,{dr' \over (r')^2 V(r')}} f(r)\quad.\label{relation}
\ee
Near the horizon the phase factor in (\ref{relation}) is precisely what one would need
to make $\phi(r,t_*)$ to be an in-coming wave, so that by simply requiring only regularity of $f(r)$ at the horizon
the mode becomes in-coming {\it automatically} ; in other words, working in the Eddington-Finkelstein
coordinate is equivalent to giving a preference to the in-coming modes, and the out-going modes
in $(r,t_*)$ coordinate would look {\it singular} in the Eddington-Finkelstein coordinate.
Reversing the logic, an inverse Eddington-Finkelstein coordinate where
\be
ds^2 = -r^2 V(r)dt_{**}^2 -2 dr dt_{**} +ds^2_{T}\quad.
\ee
should be useful for picking-up only out-going modes and for computing advanced response (Green's) functions.

\section{Holographic model I : Einstein plus $U(1)^2$ }

The simplest holographic model that implements the symmetry structure of the section 2 would
be a 5-dimensional Einstein gravity with negative cosmological constant plus a $U(1)_L\times U(1)_R$ gauge theory corresponding to the chiral symmetry in 4-dimensions. In addition, the 4-dimensional triangle anomaly of $U(1)_L\times U(1)_R$ will be reflected as a 5-dimensional Chern-Simons term in the holographic model.
The minimal Lagrangian is then
\bear
\left(16\pi G_5\right){\cal L} &=&
R +12 -{1\over 4}(F_L)_{MN}(F_L)^{MN} - {1\over 4}(F_R)_{MN}(F_R)^{MN}\\
&+& {\kappa\over 4 \sqrt{-g_5}}\epsilon^{MNPQR}\Bigg((A_L)_M (F_L)_{NP} (F_L)_{QR} -
(A_R)_M (F_R)_{NP} (F_R)_{QR}\Bigg)\quad,\nonumber
\eear
where we normalized the cosmological constant to have a unit radius for $AdS_5$ for simplicity.
The capital letters $M,N,\ldots$ denote 5-dimensional indices while Greek letters $\mu,\nu,\ldots$
will be reserved for 4-dimensional coordinates. The epsilon symbol in the above is purely numerical as we factor out $\sqrt{-g_5}$ explicitly, and our convention is $\epsilon^{r\mu\nu\alpha\beta}=\epsilon^{\mu\nu\alpha\beta}$.

It is not difficult to relate the value of $\kappa$ with the 4-dimensional anomaly coefficient \cite{Son:2009tf}.
The equation of motion for $A_L$ is
\be
\nabla_N(F_L)^{MN} - {3\kappa\over 4\sqrt{-g_5}}\epsilon^{MNPQR}(F_L)_{NP}(F_L)_{QR}=0\quad,
\ee
and in the pure $AdS_5$ vacuum in Poincare coordinate
\be
ds^2 = -r^2 dt_*^2 +{dr^2\over r^2} + r^2 \sum_{i=1}^3 (dx^i)^2\quad,
\ee
the $M=r$ component becomes
\be
\partial_\mu (F_L)^{r\mu}
- {3\kappa\over 4 r^3}\epsilon^{\mu\nu\alpha\beta}(F_L)_{\mu\nu}(F_L)_{\alpha\beta}=0\quad.\label{eqL}
\ee
According to AdS/CFT dictionary, the 4-dimensional $U(1)_L$ current that $A_L$ corresponds to
is given by
\be
j_L^\mu = -{1\over 16\pi G_5} \lim _{r\to\infty} r^3 (F_L)^{r\mu}\quad,
\ee
while the external weakly gauging potential that couples to $j_L^\mu$ is the boundary value of $A_L(r=\infty)$, so that by taking $r\to\infty$ limit of (\ref{eqL}), we have
\be
\partial_\mu j_L^\mu = -{3\kappa\over 64 \pi G_5} \epsilon^{\mu\nu\alpha\beta}(F_L)_{\mu\nu}(F_L)_{\alpha\beta}\quad,
\ee
where $F_{L}$ now represents the 4-dimensional weakly gauging potential coupled to $j_L$.
In comparison to (\ref{anomaly}), we thus have
\be
\kappa=-{2 G_5\over 3\pi }\left(N^{eff}_F N_c\right)\quad.\label{coef}
\ee

Because we are going to turn on the axial chemical potential, it is more convenient to work in terms of
$A_{em}$ and $A_a$ related to $A_{L,R}$ by (\ref{trans}),
\be
A_L= e A_{em} -A_a\quad,\quad
A_R= e A_{em} +A_a \quad,
\ee
where these fields now represent 5-dimensional gauge fields accordingly.
The action density then takes a form
\bear
\left(16\pi G_5\right){\cal L} &=&
R +12 -{e^2\over 2}(F_{em})_{MN}(F_{em})^{MN} - {1\over 2}(F_a)_{MN}(F_a)^{MN}\\
&-& {\kappa\over 2 \sqrt{-g_5}}\epsilon^{MNPQR}\Bigg(3e^2(A_a)_M (F_{em})_{NP} (F_{em})_{QR} +
(A_a)_M (F_a)_{NP} (F_a)_{QR}\Bigg)\quad,\nonumber
\eear
whose equations of motion are
\bear
&&R_{MN} +\left(4+{e^2\over 6}\left(F_{em}\right)^2 +{1\over 6}\left(F_a\right)^2 \right)g_{MN}
-e^2 \left(F_{em}\right)_{PM} \left(F_{em}\right)^P_{\,\,\,\,N}-
\left(F_{a}\right)_{PM} \left(F_{a}\right)^P_{\,\,\,\,N}= 0 \,,\nonumber\\
&&\nabla_N(F_a)^{MN} +{3\kappa\over 4 \sqrt{-g_5}}\epsilon^{MNPQR}
\left(e^2 (F_{em})_{NP} (F_{em})_{QR} + (F_{a})_{NP} (F_{a})_{QR}\right) =0 \,,\nonumber\\
&&\nabla_N(F_{em})^{MN} +{3\kappa\over 2 \sqrt{-g_5}}\epsilon^{MNPQR}
(F_{a})_{NP} (F_{em})_{QR}  =0 \,.\label{eom}
\eear
The normalization of the $A_{em}$ kinetic term looks a little unconventional, but
in this way the UV boundary value of $A_{em}$ couples to the EM current $j_{em}=e(j_L+j_R)$ with
unit strength, and it is more convenient in this sense.

There is an exact AdS Reisner-Nordstrom type black-hole solution to the above equations of motion with only an axial charge being turned on, and we will use this space-time as a background representing a finite temperature plasma with an axial/chiral chemical potential. The explicit form of the solution
in Eddington-Finkelstein coordinate is
\bear
ds^2 &=& -r^2V(r) dt^2 + 2 dr dt +r^2 \sum_{i=1}^3 (dx^i)^2\quad,\nonumber\\
A_a&=&\left({Q\over r_H^2 }-{Q\over r^2}\right) dt\quad,\quad A_{em}=0 \quad,\nonumber\\
V(r)&=& 1-{m\over r^4}+{2 Q^2\over 3 r^6}\quad,
\eear
where the location of horizon is the largest root of $V(r_H)=0$, and the axial/chiral chemical potential
can be identified as
\be
\mu_a = {Q\over r_H^2}\quad.\label{chem}
\ee
Our task is to perturb the above solution by a (small) oscillatory external EM magnetic field $B_{em}^3$ along $x^3$,
and solve the resulting linearized equations of motion of (\ref{eom}) to read off an induced
EM current $j_{em}^3$ along $x^3$ direction. Holographic renormalization tells us that $j_{em}^3$
is encoded in the near UV boundary expansion of the resulting solution of $A_{em}$ by
\be
j_{em}^3 = {e^2 \over 4\pi G_5} \lim_{r\to\infty} r^2 (A_{em})_3\quad.
\ee
Note that the factor $e^2$ is from our convention of the 5D kinetic term for $A_{em}$ in the action density.

In this linearized order,
it is easy to see that the gravity fluctuations $\delta g_{MN}$ and the fluctuation of the axial gauge field $\delta A_a$ in fact do not couple to the electromagnetic fluctuation $\delta A_{em}$, so that
the problem is much easier than naively expected. For example, in the second equation of (\ref{eom})
it is consistent to turn off $\delta A_a =0$ as $\delta A_{em}$'s contribution would be only quadratic.
The same is true for $\delta g_{MN}$ in the Einstein equation, so that for our purpose we can simply
let $\delta g_{MN}=\delta A_a=0$ and just consider the equation
\be
 \nabla_N(F_{em})^{MN} +{3\kappa\over 2 \sqrt{-g_5}}\epsilon^{MNPQR}
(F_{a})_{NP} (F_{em})_{QR}  =0\quad,
\ee
with $A_a$ being replaced by the background value in the above solution.
This decoupling between $\delta A_{em}$ and $\delta g_{MN}$ has a physical interpretation :
to linear order, external electromagnetic perturbations do not introduce additional energy-momentum,
especially there would be no momentum flow induced at first order. For example, in the microscopic picture for chiral magnetic effects in section 2, if we keep electrical neutrality by
\be
N\left(q^+_{-1/2}\right)-N\left(q^-_{+1/2}\right)= -\left(N\left(q^+_{+1/2}\right)- N\left(q^-_{-1/2}\right)\right)\quad,
\ee
the left-hand side is proportional to the chemical potential $\mu_L$ for $q_L$, and vice versa for the right-hand-side and $q_R$. Given a temperature, the thermal physics of $q_L$ would be precisely mirror
to the physics of $q_R$, so that we can expect that the number of $N\left(q^+_{-1/2}\right)$ from $q_L$
is precisely equal to $N\left(q^-_{-1/2}\right)$ from $q_R$ in the situation of $\mu_L=-\mu_R$.
Similarly, $N\left(q^-_{+1/2}\right)=N\left(q^+_{+1/2}\right)$. This gives us at the end
\be
N\left(q^+_{-1/2}\right)+N\left(q^-_{+1/2}\right)= N\left(q^+_{+1/2}\right)+ N\left(q^-_{-1/2}\right)\quad.
\ee
However, observe that the right-hand side is the total number of particles/anti-particles that move along the magnetic field, while the left-hand side is the total number that move opposite to it, so that we conclude there would be {\it no net momentum flow induced along the magnetic field}. This is the physics reason behind the decoupling between $\delta g_{MN}$ and $\delta A_{em}$, which should be universal for
any EM perturbations in neutral systems.

This feature would be absent in the set-up having only one $U(1)$ symmetry with a triangle anomaly,
as recently studied in ref.\cite{Son:2009tf}.
In this case, we have, say, $q_L$ only without $q_R$. Turning on a chemical potential for $q_L$, one
would have an excess of $q^+_{-1/2}$ over $q^-_{+1/2}$ ; $N(q^+_{-1/2})>N(q^-_{+1/2})$,
and an external magnetic field would move them together in the opposite direction to it, and would subsequently induce a net negative chiral magnetic current. This effect corresponds to $\xi_B$ in ref.\cite{Son:2009tf} for a static external magnetic field.
However, as the particles/anti-particles from $q_L$ move
together in the same direction, without having a compensating flow by $q_R$, there will be a net momentum flow introduced by the effect. This complicates choosing a right Landau frame on which there shouldn't be
any momentum flow by definition. This might explain the complicated second term of $\xi_B$ in ref.\cite{Son:2009tf}.

Back to our main purpose, the linearized equations for $A_{em}$ in components are
\bear
M=r\quad &:&\quad \nabla_N F^{rN} =0\quad,\nonumber\\
M=t\quad &:&\quad \nabla_N F^{tN}=0\quad,\nonumber\\
M=i\quad &:&\quad \nabla_N F^{iN}+{6\kappa Q\over r^6}\epsilon^{ijk} F_{jk}=0\quad,
\eear
where we omitted and will omit $\delta$ and subscript $em$ in $\delta A_{em}$  for clarity in the following. The non-vanishing Christofel symbols are
\bear
&&\Gamma^r_{rt}=\Gamma^r_{tr}=-\Gamma^t_{tt}= -{1\over 2}\partial_r\left(r^2 V(r)\right)\quad,\quad
\Gamma^r_{tt}={1\over 2} r^2 V(r) \partial_r\left(r^2 V(r)\right)\quad,\nonumber\\
&&\Gamma^i_{rj}=\Gamma^i_{jr} = {1\over r}\delta^i_j\quad,\quad\Gamma^r_{ij}=-r^3 V(r)\delta_{ij}\quad,\quad
\Gamma^t_{ij}=-r \delta_{ij}\quad,
\eear
from which one arrives at the following explicit equations
\bear
&&\left(\partial_t F_{tr}\right)+{1\over r^2}\left(\partial_i F_{ti}\right)
+V(r)\left(\partial_i F_{ri}\right) =0 \quad,\nonumber\\
&&\partial_r \left(r^3 F_{tr}\right) -r\left(\partial_i F_{ri}\right) =0\quad,\label{expeq}\\
&&\partial_r\left(r F_{ti} +r^3 V(r) F_{ri}\right)
+r\left(\partial_t F_{ri}\right) +{1\over r}\left(\partial_j F_{ji}\right)
-{6\kappa Q\over r^3} \epsilon^{ijk} F_{jk} =0\quad.\nonumber
\eear

We should fix our solution ansatz in accord to our purpose of computing time-dependent chiral magnetic conductivity. First, one can always work in the radial gauge $A_r=0$. Moreover, as we wouldn't expect
any net EM charge density appears in response to the homogeneous magnetic field, one can also adopt
the ansatz $A_t=0$. We will see the consistency of this ansatz later as it will consistently solve the full equations of motion.
We assume a definite frequency $\omega$ for every field
\be
A_i= A_i(r,x^i) e^{-i\omega t}\quad,
\ee
and we require homogeneity along the $x^3$ direction in which we turn on the magnetic field $B^3=F_{12}$,
so that we simply drop $\partial_3$ in the equations of motion. Again, the consistency of these can be checked by the full equations of motion that our solutions will solve.
As $F_{tr}=0$ in our ansatz, the first two equations in (\ref{expeq}) are uniquely solved by
\be
\partial_1 A_1+\partial_2 A_2 =0\quad.
\ee
At least we should have a non-zero $F_{12}=\partial_1 A_2-\partial_2 A_1$, so that we need to give a
finite wave-number $k_\perp=(k_1,k_2)$ to the transverse $x^\perp=(x^1,x^2)$ coordinate, although we will
take a homogeneous limit $k_\perp\to 0$ at the end to get a finite value of chiral magnetic conductivity
with only frequency dependence. These steps finally give us the following ansatz
\bear
A_1 &=& k_2 f(r) e^{-i\omega t+ik_\perp\cdot x^\perp}\quad,\nonumber\\
A_2 &=& -k_1 f(r) e^{-i\omega t+ik_\perp\cdot x^\perp}\quad,\nonumber\\
A_3 &=& g(r) e^{-i\omega t+ik_\perp\cdot x^\perp}\quad.
\eear
The remaining equation to solve, i.e. the third equation in (\ref{expeq}), gives us the following set of
two equations for $f(r)$ and $g(r)$,
\bear
&&\partial_r\left(-i\omega r f+r^3 V(r)\left(\partial_r f\right)\right) -i\omega r
\left(\partial_r f\right) -{1 \over r}k_\perp^2 f  -{i 12\kappa Q\over r^3} g =0\,,\nonumber\\
 &&
 \partial_r\left(-i\omega r g+r^3 V(r)\left(\partial_r g\right)\right) -i\omega r
\left(\partial_r g\right) -{1 \over r}k_\perp^2 g  +{i 12\kappa Q\over r^3}k_\perp^2  f =0\,,\label{modeeq}
\eear
With these being solved, the full equations of motion are satisfied within our ansatz.

It is important to determine the right boundary conditions for the radial profiles $f(r)$ and $g(r)$.
As discussed in section 3, {\it simple regularity} at the horizon $r=r_H$ will be enough to
implement the in-coming boundary condition that is suitable for a causal retarded response.
At the UV boundary $r\to\infty$, we need to have an {\it external} magnetic field $B^3=F_{12}$. Because of
\be
F_{12}=-i k_\perp^2 f(r) e^{-i\omega t+ik_\perp\cdot x^\perp}\quad,
\ee
this dictates that $f(\infty)\neq 0$. On the other hand, we shouldn't get any external perturbation
from $A_3$, which imposes the normalizable boundary condition on $g(r)$,
\be
g(r)\to {\cal O}\left({1\over r^2}\right)\quad,\quad r\to\infty\quad.\label{gbdry}
\ee
In fact, the induced EM current along $x^3$ will be obtained by
\be
j^3_{em} = {e^2\over 4\pi G_5} \lim_{r\to\infty} r^2 g(r) e^{-i\omega t+ik_\perp\cdot x^\perp}\quad,
\ee
so that our desired chiral magnetic conductivity will be computed as
\be
\sigma(\omega,k_\perp) = {i e^2\over 4\pi G_5} \lim_{r\to\infty} {r^2 g(r)\over k_\perp^2 f(r)}\quad.
\ee
One can easily convince oneself that the equations (\ref{modeeq}) with the above boundary conditions
pose a well-defined procedure of calculating $\sigma(\omega,k_\perp)$.
One should resort to numerical jobs to proceed further, however.

Focusing on a special case of homogeneity limit $k_\perp \to 0$, one indeed gets a finite value of chiral magnetic conductivity. Expanding $f(r)$ in power series of $k_\perp^2$,
\be
f=f_0 + k_\perp^2 f_1 +\cdots\quad,
\ee
one finds that $g(r)$ should start its expansion from ${\cal O}(k_\perp^2)$ for consistency,
\be
g(r)=k_\perp^2 g_1 +\cdots\quad.
\ee
This is because the {\it homogeneous} differential equation
\be
\partial_r\left(-i\omega r \cdot+r^3 V(r)\left(\partial_r \cdot \right)\right) -i\omega r
\left(\partial_r \cdot\right)=0\quad,
\ee
that $g_0$ would have to satisfy has a unique solution regular at $r=r_H$ up to an overall multiplication
factor, and its UV boundary value at $r\to\infty$ is in general finite unless one encounters a quasi-normal mode, but this happens only with a complex $\omega$ with a negative imaginary part. However, this would
contradict to our boundary condition (\ref{gbdry}), and one necessarily has $g_0=0$.

\begin{figure}[t]
\begin{center}
\includegraphics[width=7cm,height=7cm]{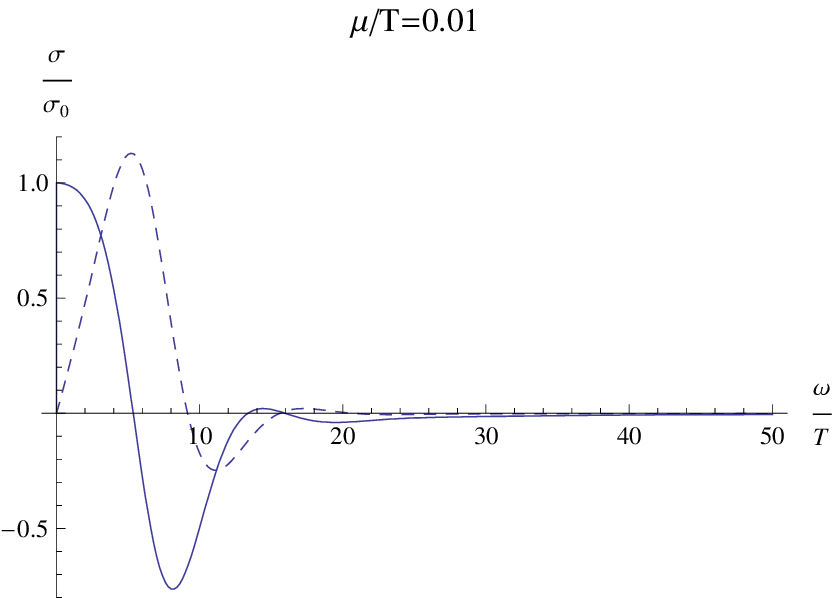}\includegraphics[width=7cm,height=7cm]{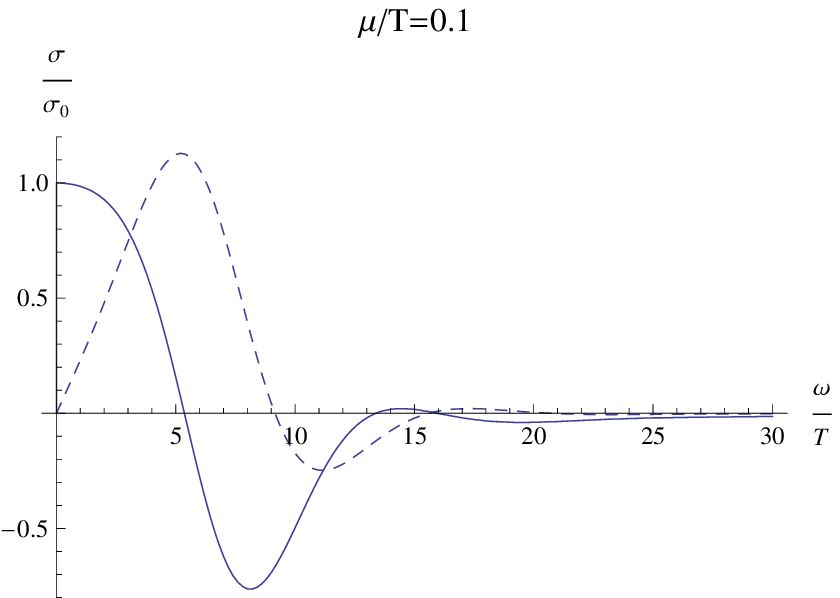}
\caption{\small Frequency dependent chiral magnetic conductivity $\sigma(\omega)$ for various axial chemical potentials. The solid line is the real part of $\sigma(\omega)$ while the dashed one is the imaginary part. }
\label{first}
\end{center}
\end{figure}

\begin{figure}[t]
\begin{center}
\includegraphics[width=7cm,height=7cm]{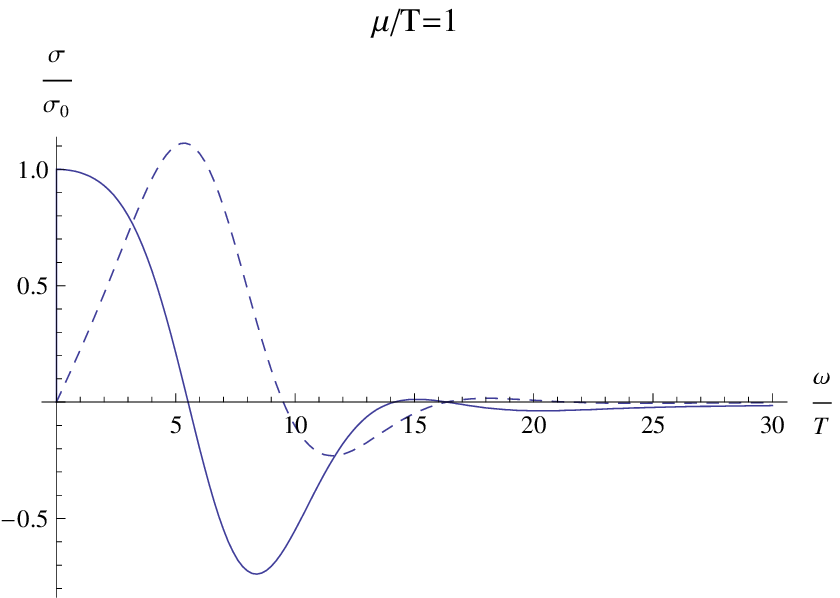}\includegraphics[width=7cm,height=7cm]{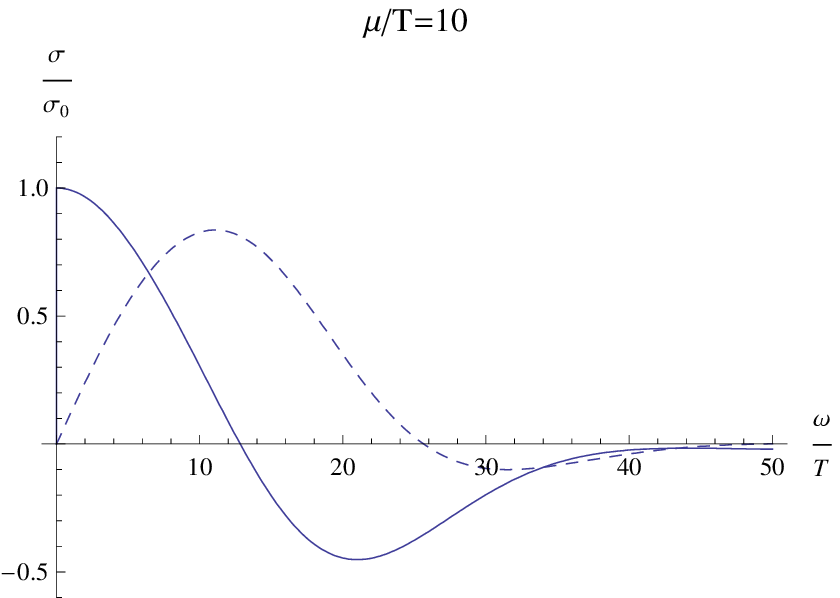}
\caption{\small More results for other values of axial chemical potentials.}
\label{first2}
\end{center}
\end{figure}

Inserting the above series expansions into (\ref{modeeq}), one gets
\bear
&&
\partial_r\left(-i\omega r f_0 +r^3 V(r) \partial_r f_0 \right) -i\omega r \left(\partial_r f_0\right) =0\quad,\nonumber\\
&&\partial_r\left(-i\omega r g_1 +r^3 V(r) \partial_r g_1 \right) -i\omega r \left(\partial_r g_1\right)
+{i 12 \kappa Q\over r^3} f_0 =0\quad,\label{homomode}
\eear
which is enough to obtain the homogeneous limit of chiral magnetic conductivity
\be
\sigma(\omega)=\sigma(\omega,k_\perp\to 0)={i e^2\over 4\pi G_5} \lim_{r\to\infty} {r^2 g_1(r)\over f_0(r)}\quad.\label{homoge}
\ee
Numerical job for computing $\sigma(\omega)$ with (\ref{homomode}) is much simpler than the general case of $\sigma(\omega,k_\perp)$ : One can first solve $f_0$ from the first equation, and then use this to obtain $g_1$ subsequently. See Figure \ref{first} and Figure \ref{first2} for the numerical results of $\sigma(\omega)$ for several different axial chemical potentials. In the Appendix, we outline an easy method of producing our numerical results.

{\it Zero frequency limit $\omega \to 0$}

It is useful to check our computation for the static limit $\omega\to 0$ as it has been argued that
the static limit is universally dictated by anomaly irrespective to the details of the dynamics.
The equation for $f_0$ simplifies to
\be
\partial_r\left(r^3 V(r) \partial_r f_0\right)=0\quad,
\ee
whose unique regular solution at $r=r_H$ up to normalization is a constant : $f_0=1$.
The subsequent equation for $g_1$
\be
 \partial_r\left(r^3 V(r) \partial_r g_1\right) + {i 12 \kappa Q\over r^3}=0\quad,
\ee
with the regularity at $r=r_H$ as well as the normalizability (\ref{gbdry}) at UV determines the
unique solution
\be
g_1(r)=i 6\kappa Q \int^r_\infty dr'\,{1\over (r')^3 V(r')}\left({1\over (r')^2} - {1\over r_H^2}\right)\quad.
\ee
Using the explicit form of $V(r)$, we in fact only need the fact that $V(r)\to 1$ as $r\to\infty$ to
find the near boundary behavior of $g_1$ as
\be
g_1(r) \to {i 3\kappa Q\over r_H^2} {1\over r^2}+\cdots\quad,\quad r\to\infty\quad,
\ee
so that the chiral magnetic conductivity from (\ref{homoge}) at zero frequency is
\be
\sigma(\omega\to 0)\equiv \sigma_0= -{3\kappa Q e^2 \over 4\pi G_5 r_H^2} ={e^2 \mu_a \over 2\pi^2}\left(N^{eff}_F N_c\right)\quad,\label{zerofre}
\ee
where we used (\ref{coef})
\be
\kappa=-{2 G_5\over 3\pi }\left(N^{eff}_F N_c\right)\quad,
\ee
as well as the expression for the axial/chiral chemical potential (\ref{chem})
\be
\mu_a={Q\over r_H^2}\quad,
\ee
in the last equality. This is indeed the right answer.

\section{Holographic model II : The model of Sakai and Sugimoto}

\begin{figure}[t]
\begin{center}
\includegraphics[width=12cm,height=7cm]{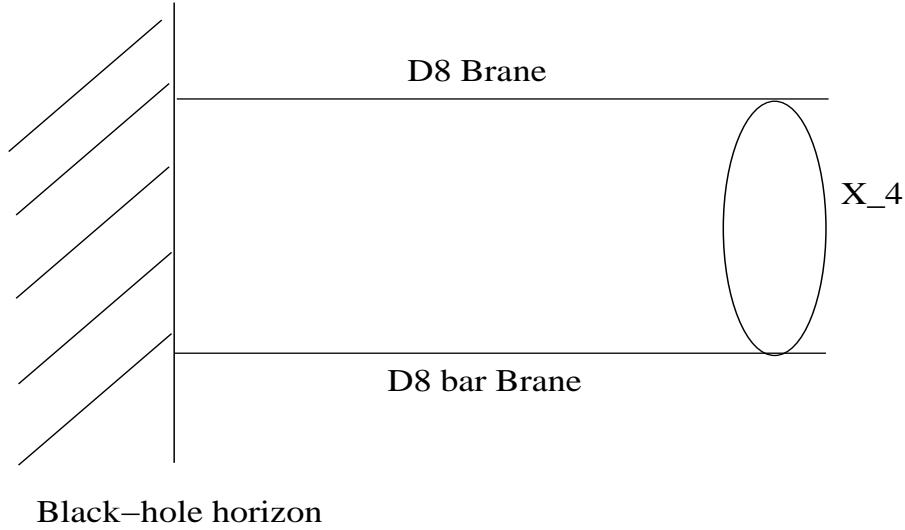}
\caption{\small A schematic picture of the Sakai-Sugimoto model in its deconfined and chiral symmetry restored phase. }
\label{d8d8bar}
\end{center}
\end{figure}

Our second model for a holographic calculation of time-dependent chiral magnetic conductivity
is the deconfined and chiral symmetry restored phase of the Sakai-Sugimoto model at finite temperature.
This model is supposed to describe, at least qualitatively, the large $N_c$ limit of QCD with a small
number of fundamental quarks in quenched approximation, and in this sense would be more realistic than
the somewhat arbitrary set-up in the previous section.
For simplicity, we will focus on a single flavor $N_F=1$ case, which means that we have a single
$D8$ and $\overline{D8}$ probe brane pair embedded into a known black-hole solution for a deconfined phase \cite{Aharony:2006da}.
The chiral symmetry $U(1)_L$ and  $U(1)_R$ that we are interested in live on the world-volumes of these
probe $D8$ and $\overline{D8}$ branes respectively. For the chiral symmetry to be restored/unbroken, we confine ourselves to the phase where
each of these branes meet the black-hole horizon and do not meet with each other. See the Figure \ref{d8d8bar} for a schematic picture.
This also implies that the leading order dynamics on each 8-branes are independent of each other :
assigning $U(1)_L$ to $D8$ and vice versa for $U(1)_R$ and $\overline{D8}$, the total action will simply be
a sum of the two 8-brane world volume actions
\be
S_{tot}= S_{D8}\left(A_L\right) + S_{\overline{D8}}\left(A_R\right) \quad,
\ee
where
\be
S_{D8/\overline{D8}} = -\mu_8 \int d^9 \xi \, e^{-\phi}\,\sqrt{{\rm det}\left(g^*+2\pi l_s^2 F\right)}
\,\,{\mp}\,\, {\mu_8\left(2\pi l_s^2\right)^3 \over 3!}\int \,F_4^{RR}\wedge A\wedge F\wedge F\quad,\label{8-act}
\ee
with $\mu_p=(2\pi)^{-p}l_s^{-(p+1)}$. Note that the Chern-Simons coupling for $\overline{D8}$ has the opposite sign to that of $D8$, and we show only the relevant term with the $F_4^{RR}$ 4-form Ramond-Ramond field strength as our background carries only this Ramond-Ramond field.

The 10-dimensional black-hole solution that provides a background corresponding to a deconfined phase
{\it in Eddington-Finkelstein coordinate} is \cite{Aharony:2006da}
\bear
ds^2&=&\left(U\over R\right)^{3\over 2}\left(-f(U)dt^2+(dx^i)^2+{1\over (M_{KK}l_s)^2}dx_4^2\right)+2 dU dt
+\left(R\over U\right)^{3\over 2}U^2  d\Omega_4^2\quad,\nonumber\\
F_4^{RR}&=&{(2\pi l_s)^3 N_c\over V_4}\epsilon_4\quad,\quad e^\phi=g_s\left(U\over R\right)^{3\over 4}\quad,
\quad V_4={\rm Vol}(S^4)={8\pi^2\over 3}\quad,\nonumber\\
R^3&=& \pi g_s N_c l_s^3\quad,\quad f(U)=1-\left(U_T\over U\right)^3\quad,\label{decon}
\eear
where the temperatute $\beta={1\over T}$ is related to the location of the black-hole horizon $U_T$ by
\be
\beta={4\pi\over  3}\left(R^3\over U_T\right)^{1\over 2}\quad,
\ee
and $x_4$ has a period $(2\pi l_s)$ which is not essential for our purpose though. The $F^{RR}_4$
is normalized in such a way that
\be
{1\over (2\pi l_s)^3} \int_{S^4} \, F^{RR}_4 = N_c\quad,
\ee
and this convention conforms to those in our writing of 8-brane actions (\ref{8-act}).
Each 8-branes of our interest spans the coordinates $(U,t,x^i,\Omega_4)$ of total 9-dimensions sitting at a point in $x_4$, but
dual QCD dynamics should be insensitive to the modes inside $S^4$, so that we will assume homogeneity
along $\Omega_4$ directions and integrate over them from the very first stage.
After that, the resulting 5-dimensional action takes a form with a minor computation
\be
S_{D8/\overline{D8}}= -C R^{9\over 4} \int dx^4 dU\,U^{1\over 4}\sqrt{{\rm det}\left(g^*+2\pi l_s^2 F\right)}
\,\mp\,{N_c\over 96\pi^2} \int dx^4 dU \,\epsilon^{MNPQR} A_M F_{NP} F_{QR}\quad,
\ee
with a definition of
\be
C={N^{1\over 2}\over 3\cdot 2^5\cdot\pi^{11\over 2} g_s^{1\over 2} l_s^{15\over 2}}\quad,
\ee
for later convenience.

Our objective is to first construct a background solution having a finite chemical potential $\mu_a$ for the axial
current
\be
A_a={1\over 2}\left(-A_L+A_L\right)\quad,
\ee
while keeping the system neutral under electromagnetism for simplicity
\be
eA_{em}={1\over 2}\left(A_L+A_R\right)\quad,
\ee
and then to consider a small external perturbation of electromagnetic magnetic field to see its retarded response in the current. This means we turn on a chemical potential $\mu_L=-\mu_a$ for the $U(1)_L$ symmetry on the $D8$ brane, and similarly $\mu_R=\mu_a$ for $U(1)_R$ on the $\overline{D8}$ brane. See ref.\cite{Aharony:2007uu,Parnachev:2007bc} for turning on iso-spin chemical potentials in the model.
To find the corresponding background solution, it is enough to keep $F_{tU}$ component only in the action
of the 8-branes, and a simple calculation gives
\be
S_{D8/\overline{D8}}=-C\int d^4 x dU\, U^{5\over 2} \sqrt{1-(2\pi l_s^2)^2 (F_{tU})^2}\quad,
\ee
whose solution is easily integrable to be
\be
F_{tU}= {\pm\alpha \over\sqrt{U^5+(2\pi l_s^2)^2\alpha^2}}\quad,
\ee
with an integration constant $\alpha >0$, where the upper sign is for $A_L$ on the $D8$ and vice versa for $A_R$ on the $\overline{D8}$. One can always work in the radial gauge $A_U=0$ in which the above solutions
can be further integrated for $A_t$ as
\be
A_t(U) = \mp \int_{U_T}^U dU'\,{\alpha \over\sqrt{(U')^5 +(2\pi l_s^2)^2\alpha^2}}\quad,
\ee
where we have imposed the condition $A_t(U=U_T)=0$ at the horizon to fix an integration constant.
The chemical potentials of $U(1)_{L,R}$ are then read off as the UV boundary values of the above $A_t$ respectively, which
relates $\alpha$ with $\mu_a$ as
\be
\mu_a = \int_{U_T}^\infty
dU'\,{\alpha \over\sqrt{(U')^5 +(2\pi l_s^2)^2\alpha^2}}={2\alpha\over 3 U_T^{3\over 2}} \,\, {_2F_1}\left({3\over 10},{1\over 2},{13\over 10},-{(2\pi l_s^2)^2\alpha^2\over U_T^5}\right)\quad.
\ee
We should remember this relation to identify a physical meaning of $\alpha$.

Having obtained the background solution, one next needs to find a linearized equation of motion for
the vector part electromagnetism treated as a perturbation to the background solution.
One way of performing the analysis is to expand the 8-brane actions up to second order in fluctuations
so that the linearized equation of motion can be derived directly from it.
Using the series expansion
\be
\sqrt{\rm det \left(1+\delta A\right)}= 1+{1\over 2}{\rm tr}\left(\delta A\right)
+{1\over 8}\left[{\rm tr}\left(\delta A\right)\right]^2 -{1\over 4}\left[{\rm tr}\left(\left(\delta A\right)^2\right)\right]+{\cal O}\left(\left(\delta A\right)^3\right)\quad,
\ee
a straightforward but substantial amount of algebra finally produces the result
\bear
S_{D8/\overline{D8}}^{(2)}&=&
C\int d^4x dU\,\Bigg[L(U)+{1\over 2}A(U)\left(\delta F_{tU}\right)^2
-B(U)\left(\delta F_{ti}\right)\left(\delta F_{Ui}\right)
-{1\over 2} C(U)\left(\delta F_{Ui}\right)^2 \nonumber\\
&-&{1\over 4} D(U)\left(\delta F_{ij}\right)\left(\delta F_{ij}\right) \Bigg]
-{N_c\over 8\pi^2}\int d^4x dU\, \epsilon^{ijk} A_t(U)\left(\delta F_{Ui}\right)\left(\delta F_{jk}\right)\quad,\label{secondaction}
\eear
where here $A_t(U)$ in the last term is
\be
A_t(U) = + \int_{U_T}^U dU'\,{\alpha \over\sqrt{(U')^5 +(2\pi l_s^2)^2\alpha^2}}\quad,
\ee
for {\it both $D8$ and $\overline{D8}$ branes}, because the sign difference from the Chern-Simons terms
is compensated by having the opposite background chemical potentials, so that the fluctuation actions for the 8-branes are now identical to each other. The appropriate functions that appear in the above are
\bear
L(U)&=& -U^{5\over 2}\left(1-(2\pi l_s^2)^2\left(F_{tU}\right)^2\right)^{1\over 2} = -U^5 \left(U^5+(2\pi l_s^2)^2\alpha^2\right)^{-{1\over 2}}\quad,\\
A(U)&=&(2\pi l_s^2)^2U^{5\over 2}\left(1-(2\pi l_s^2)^2\left(F_{tU}\right)^2\right)^{-{3\over 2}}=(2\pi l_s^2)^2 U^{-5}\left(U^5+(2\pi l_s^2)^2\alpha^2\right)^{3\over 2}\quad,\nonumber\\
B(U)&=&(2\pi l_s^2)^2\left(R\over U\right)^{3\over 2} U^{5\over 2}\left(1-(2\pi l_s^2)^2\left(F_{tU}\right)^2\right)^{-{1\over 2}}=
(2\pi l_s^2)^2\left(R\over U\right)^{3\over 2}\left(U^5+(2\pi l_s^2)^2\alpha^2\right)^{1\over 2}\quad,
\nonumber\\
C(U)&=& (2\pi l_s^2)^2 f(U) U^{5\over 2} \left(1-(2\pi l_s^2)^2\left(F_{tU}\right)^2\right)^{-{1\over 2}}=
(2\pi l_s^2)^2 f(U)\left(U^5+(2\pi l_s^2)^2\alpha^2\right)^{1\over 2}\quad,\nonumber\\
D(U)&=& (2\pi l_s^2)^2\left(R\over U\right)^{3} U^{5\over 2}\left(1-(2\pi l_s^2)^2\left(F_{tU}\right)^2\right)^{{1\over 2}}=(2\pi l_s^2)^2\left(R\over U\right)^3 U^5 \left(U^5+(2\pi l_s^2)^2\alpha^2\right)^{-{1\over 2}}\,.\nonumber
\eear
For a qualitative understanding, the detailed form of the above functions are not essential except the fact that $C(U)\to 0$ at the horizon $U\to U_T$, and their near-boundary behaviors
\bear
A(U)&\to& (2\pi l_s^2)^2 \,U^{5\over 2}\quad,\quad B(U)\to (2\pi l_s^2)^2R^{3\over 2}\, U\quad,\nonumber\\
C(U)&\to& (2\pi l_s^2)^2\,U^{5\over 2}\quad,\quad D(U)\to (2\pi l_s^2)^2 R^3 \,U^{-{1\over 2}}\quad,\quad U\to\infty\quad.\label{UV}
\eear
The structure of the above fluctuation action is in fact qualitatively identical to that in our first holographic model, which will be more manifest later.

As the $D8$-brane action for fluctuations up to second order has an identical form to that of the $\overline{D8}$-brane, the total action for the electromagnetic fluctuations would be
simply a {\it twice} of (\ref{secondaction}) with the replacement
\be
\delta F_{MN} \to e (\delta F_{em})_{MN}\quad,
\ee
where we will omit the subscript $em$ from now on for clarity, while we will leave $\delta$.
From this action with the near-boundary behaviors (\ref{UV}), a gauge/gravity dictionary for the
current can be easily deduced as
\be
j_{em}^3 = 3 e^2 C (2\pi l_s^2)^2 \lim_{U\to\infty} U^{3\over 2} \left(\delta A_3\right)\quad,\label{holoRG}
\ee
which we will use later to find the induced electromagnetic current from the solution of $\delta A$
in response to an external EM magnetic field.

It is straightforward to write down the equations of motion from (\ref{secondaction}),
\bear
&&\partial_U\left(A(U)\delta F_{tU}\right) -B(U)\left(\partial_i \delta F_{Ui}\right)=0\quad,\\
&&A(U)\left(\partial_t \delta F_{tU}\right)+B(U) \left(\partial_i \delta F_{ti}\right) +C(U)\left(\partial_i \delta F_{Ui}\right)=0\quad,\nonumber\\
&& B(U)\left(\partial_t\delta F_{Ui}\right) + \partial_U\left(B(U)\delta F_{ti}+C(U)\delta F_{Ui}\right)
+D(U) \partial_j\left(\delta F_{ji}\right)-{N_c\over 8\pi^2 C} F_{tU}\epsilon^{ijk}\delta F_{jk}=0\,,\nonumber
\eear
whose structure is essentially similar to (\ref{expeq}) in comparison as it should be expected.
Subsequent analysis henceforth is quite close to the previous holographic model.
Starting from the ansatz
\bear
\delta A_1 &=& k_2 f(U) e^{-i\omega t+ik_\perp\cdot x^\perp}\quad,\nonumber\\
\delta A_2 &=& -k_1 f(U) e^{-i\omega t+ik_\perp\cdot x^\perp}\quad,\nonumber\\
\delta A_3 &=& g(U) e^{-i\omega t+ik_\perp\cdot x^\perp}\quad.
\eear
the full equations of motion are solved by the following two equations for $f(U)$ and $g(U)$,
\bear
&&\partial_U\left(C(U)\left(\partial_U f\right)-i\omega B(U) f\right)-i\omega B(U) \left(\partial_U f\right)
-D(U) k_\perp^2 f  -{i N_c\over 4\pi^2 C} F_{tU}(U) g =0\quad,\nonumber\\
&&\partial_U\left(C(U)\left(\partial_U g\right)-i\omega B(U) g\right)-i\omega B(U) \left(\partial_U g\right)
-D(U) k_\perp^2 g  +{i N_c\over 4\pi^2 C} F_{tU}(U) k_\perp^2 f =0\quad,\nonumber\\
\eear
with the boundary conditions that they are regular at the horizon $U=U_T$ and $g(U)$
should be normalizable at UV,
\be
g(U)\sim {1\over U^{3\over 2}}\quad,\quad U\to\infty\quad.
\ee
Using the current formula (\ref{holoRG}), and the external magnetic field
\be
\delta F_{12}=-i k_\perp^2 f(\infty) e^{-i\omega t+ik_\perp\cdot x^\perp}\quad,
\ee
the momentum dependent chiral magnetic conductivity will be computed as
\be
\sigma(\omega,k_\perp)= i 3 e^2 C (2\pi l_s^2)^2\lim_{U\to\infty} {U^{3\over 2} g(U)\over k_\perp^2 f(U)}\quad.
\ee
For
the homogeneous limit $k_\perp\to 0$, one expands
\be
f=f_0+k_\perp^2 f_1+\cdots\quad,\quad g=k_\perp^2 g_1+\cdots\quad,
\ee
and $f_0$ and $g_1$ are solved by the equations
\bear
&&\partial_U\left(C(U)\left(\partial_U f_0\right)-i\omega B(U) f_0\right)-i\omega B(U) \left(\partial_U f_0\right)
=0\quad,\\
&&\partial_U\left(C(U)\left(\partial_U g_1\right)-i\omega B(U) g_1\right)-i\omega B(U) \left(\partial_U g_1\right)
+{i N_c\over 4\pi^2 C} F_{tU}(U)f_0 =0\quad,\nonumber
\eear
where one first solves the former equation and then use that to solve the second equation.
The homogeneous but frequency dependent chiral magnetic conductivity is then
\be
\sigma(\omega) = i 3 e^2 C (2\pi l_s^2)^2\lim_{U\to\infty} {U^{3\over 2} g_1(U)\over  f_0(U)}\quad.\label{2result}
\ee

{\it Static limit $\omega\to 0$ }

We check the static limit of our formula for consistency.
In $\omega\to 0$ limit, the equation for $f_0$ is
\be
\partial_U\left(C(U)\partial_U f_0\right)=0\quad,
\ee
whose regular solution at the horizon is simply a constant, $f_0=1$, due to the fact $C(U_T)=0$.
The equation for $g_1$ is then
\be
\partial_U\left(C(U)\partial_U g_1\right) = -{i N_c\over 4\pi^2 C} F_{tU}=\partial_U\left({i N_c\over 4\pi^2 C} A_t(U)\right)\quad,
\ee
whose integration is
\be
\partial_U g_1 ={i N_c\over 4\pi^2 C}{A_t(U)\over C(U)}\quad,
\ee
where we have used the fact $A_t(U_T)=0$ so that the right-hand side is regular at $U=U_T$.
Integrating once more, one gets
\be
g_1(U)={i N_c\over 4\pi^2 C}\int_{\infty}^U dU'\,{A_t(U')\over C(U')}\quad,
\ee
where we already chose an integration constant to have a normalizable solution for $g_1$.
From the UV asymptotic behavior of $C(U)$
\be
C(U)\to (2\pi l_s^2)^2 U^{5\over 2}\quad, \quad U\to\infty\quad,
\ee
and also the previous identification of the chemical potential
\be
A_t(\infty)=\mu_a\quad,
\ee
one finally arrives at
\be
g_1(U)\to -{i N_c\over 4\pi^2 C}{2\over 3} {\mu_a\over (2\pi l_s^2)^2 }{1\over U^{3\over 2}}+\cdots\quad,\quad U\to\infty\quad.
\ee
Inserting into (\ref{2result}), one checks that
\be
\sigma(\omega=0)\equiv \sigma_0= {e^2 \mu_a\over 2\pi^2} \cdot N_c\quad,\label{zero2}
\ee
which is the right result for a single flavor $N_F=1$ quark we are considering.
\begin{figure}[t]
\begin{center}
\includegraphics[width=7cm,height=7cm]{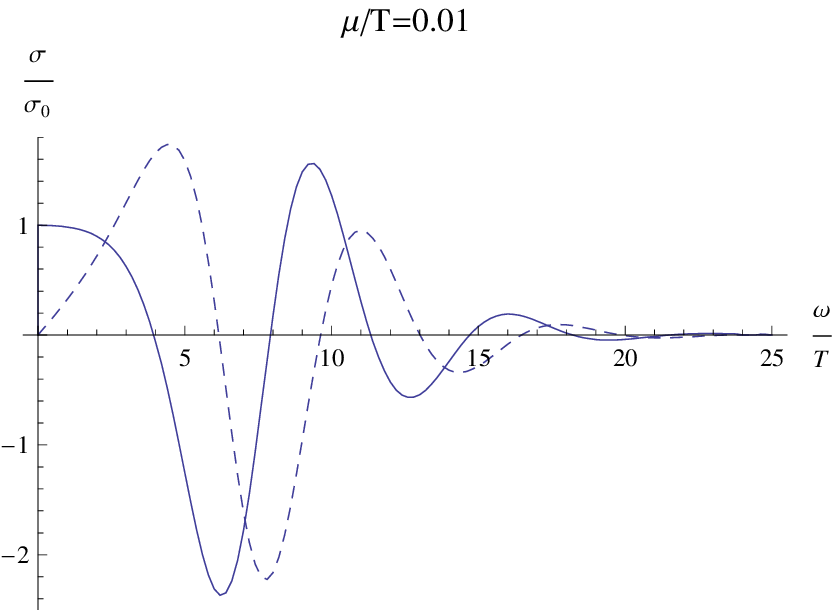}\includegraphics[width=7cm,height=7cm]{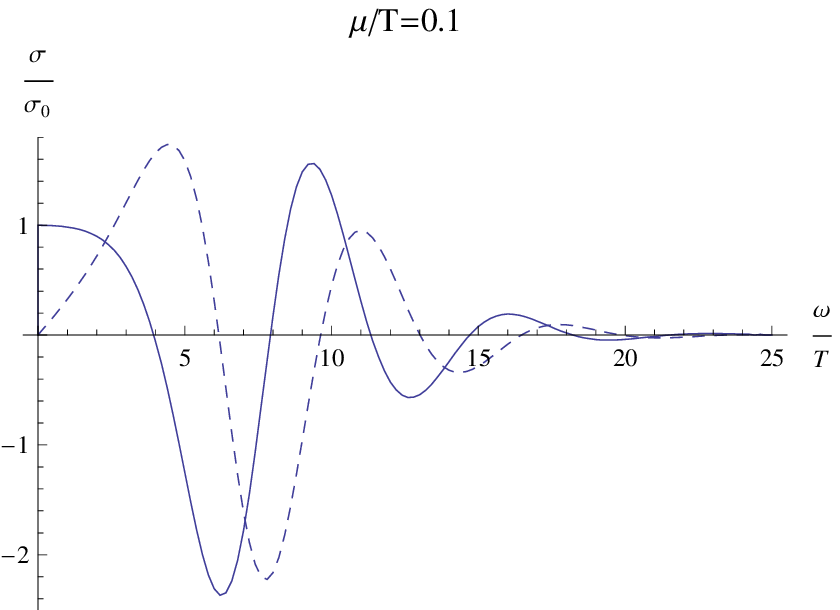}
\caption{\small Time-dependent chiral magnetic conductivity $\sigma(\omega)$ for various axial chemical potentials in the Sakai-Sugimoto model with $T=200$ MeV. The solid line is the real part of $\sigma(\omega)$ while the dashed one is the imaginary part. }
\label{second}
\end{center}
\end{figure}
\begin{figure}[t]
\begin{center}
\includegraphics[width=7cm,height=7cm]{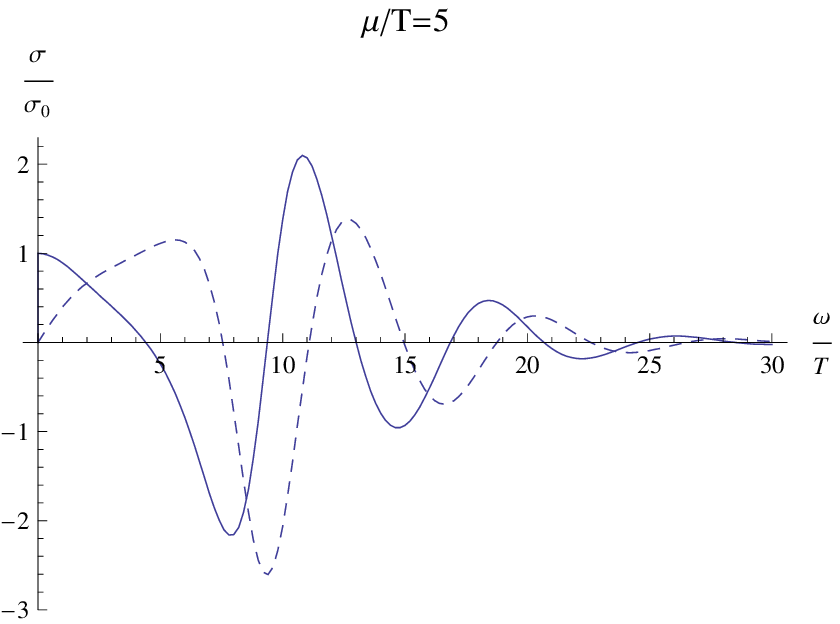}\includegraphics[width=7cm,height=7cm]{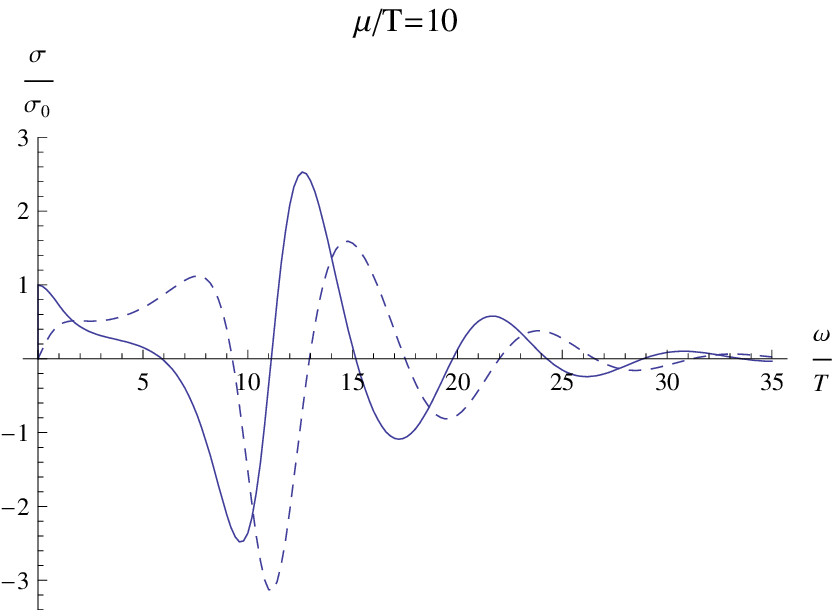}
\caption{\small More results in the Sakai-Sugimoto model with $T=200$ MeV for other values of axial chemical potentials.}
\label{second2}
\end{center}
\end{figure}

To perform numeric jobs in the case of Sakai-Sugimoto model, one has to fix the parameters of the theory. First one can always work in the unit where $2\pi l_s^2=1$. From the $\rho$ meson mass and the pion decay constant, Sakai and Sugimoto determined the parameters \cite{Sakai:2004cn,Sakai:2005yt}
\be
g_{YM}^2 N_c \approx 17\quad,\quad M_{KK}=0.94\,{\rm GeV}\quad,
\ee
and we will take these values for illustrative purposes. For the temperature, we choose $T=200$ MeV
as a representative value relevant for the RHIC experiment\footnote{According to ref.\cite{Aharony:2006da}, the confinement/deconfinement transition happens at $T_c={M_{KK}\over 2\pi}\approx 150$ MeV in this model. See also ref.\cite{Peeters:2007ab} for further aspects.}. With these being fixed, there is no other ambiguity in the model.
We present the numerical results of $\sigma(\omega)$ in Figure \ref{second} and Figure \ref{second2}.

\section{Summary of results}

We set-up a couple of holographic frameworks for computing time-dependent chiral magnetic conductivity.
In the first model, we consider a full back-reacted Reisner-Nordstrom black-hole solution with only an axial chemical potential turned on, to study the induced electromagnetic current in response to a small, time-dependent magnetic field perturbation. Our second model is based on the more realistic model of Sakai and Sugimoto in its deconfined and chiral symmetry restored phase, but within a quenched/probe approximation.
Both models give us qualitatively similar results for the frequency dependent chiral magnetic conductivity, which may be a useful complementary computation for the strongly coupled regime to the existing recent weak-coupling computation in perturbative QCD \cite{Kharzeev:2009pj}. Our numerical results are presented in Figures \ref{first},\ref{first2},\ref{second}, and \ref{second2} for an illustrative purpose.
As the results show, the real part of chiral magnetic conductivity stays to the value at $\omega=0$ for small $\omega$, contrary to the result in weak-coupling where it drops to $1\over 3$ as soon as $\omega\ne 0$ \cite{Kharzeev:2009pj}. As was already pointed out in ref.\cite{Kharzeev:2009pj}, it might be due to the strong interactions of the charge carriers. Consequently, the current response and charge asymmetry in RHIC plasma would be bigger than the weak-coupling result in ref.\cite{Kharzeev:2009pj}.
As a future direction,
it might be worthwhile to generalize our calculations to the cases with non-zero electromagnetic charge density.

\vskip 1cm \centerline{\large \bf Acknowledgement} \vskip 0.5cm

The author benefited much from discussions with Ki-Myeong Lee and Keun-Young Kim, and thanks Dmitri Kharzeev and Harmen Warringa for very helpful inputs and correspondences.
He also acknowledges kind hospitalities from Deog-Ki Hong and Piljin Yi while his visit to APCTP focus program and KIAS. Part of this work was done during the
APCTP program FP-02 2009. Finally, we appreciate correspondence from Rebhan, Schmitt, and Stricker.

\appendix

\section{Appendix : An easy method of performing numerical jobs}

We explain our way of solving the equation (\ref{homomode})
\bear
&&
\partial_r\left(-i\omega r f_0 +r^3 V(r) \partial_r f_0 \right) -i\omega r \left(\partial_r f_0\right) =0\quad,\label{app1}\\
&&\partial_r\left(-i\omega r g_1 +r^3 V(r) \partial_r g_1 \right) -i\omega r \left(\partial_r g_1\right)
+{i 12 \kappa Q\over r^3} f_0 =0\quad,\label{app2}
\eear
to compute $\sigma(\omega)$ numerically.
One first notices that it is straightforward to solve the equation (\ref{app1}) for $f_0$ numerically without any shooting ambiguity. For regularity at $r=r_H$ where $V(r)=0$, one simply considers $r\to r_H$ limit of (\ref{app1}) to deduce the regularity condition
\be
\left(\partial_r f_0\right)(r_H) = \left({ i\omega\over r_H^3 V'(r_H)-2 i \omega r_H} \right)f_0(r_H)\quad,
\ee
and since a normalization of $f_0$ is not important, we can start at $r=r_H$ with $f_0(r_H)=1$ and $\left(\partial_r f_0\right)(r_H)$ given in the above to solve (\ref{app1}) numerically for $r\ge r_H$.

Once $f_0$ is found, there is a nice method of solving (\ref{app2}) for $g_1$ without further numerics.
The idea is to transform the differential operator acting on $g_1$ into an {\it integrable form},
\be
\partial_r\left(-i\omega r \cdot +r^3 V(r) \partial_r \cdot \right) -i\omega r \left(\partial_r \cdot \right)= P(r)\partial_r\Bigg(R(r)\partial_r\Big(S(r)\cdot\Big)\Bigg)\quad,\label{diffop}
\ee
and comparing the both sides gives us the equations for the yet unknown functions $P(r)$, $R(r)$, and $S(r)$ as
\bear
PRS &=& r^3 V(r)\quad,\label{a1}\\
P\Bigg(R\Big(\partial_r S\Big)+\partial_r\Big(RS\Big)\Bigg) &=& \partial_r\left(r^3 V(r)\right)-2i\omega r\quad,\label{a2}\\
P\partial_r\Bigg(R\Big(\partial_r S\Big)\Bigg) &=& -i\omega \label{a3} \quad.
\eear
Replacing $R$ in the second equation (\ref{a2}) by using (\ref{a1}), one gets a nice simplification
\be
\partial_r\,{\rm ln}\left(S\over P\right) = -{2 i\omega \over r^2 V(r)}\quad,
\ee
whose integration gives
\be
P^{-1}(r)=S^{-1}(r) \,e^{-2i\omega \int^r_{\infty} {dr'\over (r')^2 V(r')}}\quad.
\ee
Using this and (\ref{a1}) to replace $P$ and $R$ in the last equation (\ref{a3}), one gets a second order differential equation for $S^{-1}$, which turns out to be precisely the original homogeneous
equation with the differential operator (\ref{diffop}), whose {\it regular} solution we already obtained
as $f_0$ numerically. Therefore, one can simply let $S^{-1}=f_0$ and one finally has
\bear
S^{-1}(r)&=&f_0(r)\quad,\nonumber\\
P^{-1}(r)&=&f_0(r) \,e^{-2i\omega \int^r_{\infty} {dr'\over (r')^2 V(r')}}\quad,\nonumber\\
R(r)&=& r^3 V(r)\left(f_0(r)\right)^2 \,e^{-2i\omega \int^r_{\infty} {dr'\over (r')^2 V(r')}}\quad.\label{prs}
\eear

The equation (\ref{app2}) for $g_1$ subsequently takes a form
\be
P(r)\partial_r\Bigg(R(r)\partial_r\Big(S(r)g_1\Big)\Bigg) = -{i 12 \kappa Q\over r^3} f_0\quad,
\ee
whose first integration results in
\be
R(r)\partial_r\Big(S(r) g_1\Big)=-i 12\kappa Q \int^r_{r_H} dr'\,{f_0(r')\over (r')^3 P(r')}\quad,
\ee
where we fix an integration constant by considering $r=r_H$ where the left-hand side vanishes by $R(r_H)=0$.  The subsequent integration gives
\be
g_1(r)=-i 12\kappa Q S^{-1}(r)\int^r_{\infty} {dr'\over R(r')} \int^{r'}_{r_H} dr''\,{f_0(r'')\over (r'')^3 P(r'')}\quad,
\ee
where the second integration constant is fixed to make $g_1$ normalizable at $r\to\infty$.
With the explicit choice of $P$, $R$ and $S$ in (\ref{prs}), one obtains $g_1$ solely in terms of the known solution $f_0$ as
\be
g_1(r)=-i 12\kappa Q f_0(r)\int_{\infty}^r dr'\,{e^{2i\omega\int^{r'}_\infty{dr''\over (r'')^2 V(r'')}}
\over (r')^3 V(r')(f_0(r'))^2} \int^{r'}_{r_H} dr''\,{(f_0(r''))^2\over (r'')^3} e^{-2i\omega \int^{r''}_\infty {dr'''\over (r''')^2 V(r''')}}\,.
\ee

Having solved for $g_1$, what one needs in order to find $\sigma(\omega)$ is the near boundary expansion of $g_1(r)$ in $r\to\infty$.
From $V(\infty)=1$ and the fact that $f_0(\infty)$ is some finite constant, it is indeed checked easily that $g_1$ has a behavior $\sim {1\over r^2}$ as $r\to\infty$, and more precisely
\be
g_1(r)\to \left({i 6 \kappa Q\over  f_0(\infty)  }\int^\infty_{r_H}dr'\,{(f_0(r'))^2\over (r')^3} e^{-2i\omega \int^{r'}_\infty {dr''\over (r'')^2 V(r'')}}\right){1\over r^2} +\cdots\quad.
\ee
In conjunction with our formula for $\sigma(\omega)$ (\ref{homoge}) and the result for the zero frequency limit $\sigma_0$ (\ref{zerofre}),
\be
\sigma(\omega\to 0)\equiv \sigma_0= -{3\kappa Q e^2 \over 4\pi G_5 r_H^2} \quad,
\ee
one finally arrives at the useful expression
\be
{\sigma(\omega)\over \sigma_0}={2 r_H^2\over \big(f_0(\infty)\big)^2}
\int^\infty_{r_H}dr'\,{(f_0(r'))^2\over (r')^3} e^{-2i\omega \int^{r'}_\infty {dr''\over (r'')^2 V(r'')}}\quad,
\ee
which can be directly computed numerically once we find $f_0$ only. Our plots are generated using this technique.

The above procedure can easily be repeated in the case of Sakai-Sigimoto model too, and we simply provide the results of each steps. For simplicity, we work in the unit $2\pi l_s^2=1$.
Once $f_0(U)$ is found numerically by requiring at the horizon $U=U_T$
\be
\left(\partial_U f_0\right)(U_T) = \left({ i\omega \left(\partial_U B\right)(U_T)\over \left(\partial_U C\right)(U_T)-2 i \omega B(U_T)} \right)f_0(U_T)\quad,
\ee
the differential operator acting on $g_1$ can be transformed to an integrable form with
\bear
S^{-1}(U)&=&f_0(U)\quad,\nonumber\\
P^{-1}(U)&=& f_0(U) \,e^{-2i\omega\int^U_{\infty} dU'\,{B(U')\over C(U')}}\quad,\nonumber\\
R(U)&=& C(U) \Big(f_0(U)\Big)^2 \,e^{-2i\omega\int^U_{\infty} dU'\,{B(U')\over C(U')}}\quad,
\eear
so that the equation for $g_1(U)$ is integrated as
\be
g_1(U)=-{iN_c\over 4\pi^2 C}f_0(U)\int^U_\infty dU'\,{e^{2i\omega\int^{U'}_\infty dU''\,{B(U'')\over C(U'')}}\over C(U')\Big(f_0(U')\Big)^2 } \int^{U'}_{U_T} dU''\, F_{tU}(U'')\Big(f_0(U'')\Big)^2
e^{-2i\omega\int^{U''}_\infty dU'''\,{B(U''')\over C(U''')}}\,,
\ee
where
\be
F_{tU}(U)=-{\alpha\over\sqrt{U^5+\alpha^2}}\quad.
\ee
From the near boundary behavior $C(U)\to U^{5\over 2}$ as $U\to\infty$, one deduces without difficulty that
\be
g_1(U)\to \left({i N_c\over 6\pi^2 C f_0(\infty)}
\int^{\infty}_{U_T} dU'\, F_{tU}(U')\Big(f_0(U')\Big)^2
e^{-2i\omega\int^{U'}_\infty dU''\,{B(U'')\over C(U'')}}\right){1\over U^{3\over 2}}+\cdots\quad,
\ee
from which in conjunction with (\ref{2result}) and (\ref{zero2}),
\be
\sigma(\omega=0)\equiv \sigma_0= {e^2 \mu_a\over 2\pi^2} \cdot N_c\quad,
\ee
one finally obtains the expression
\be
{\sigma(\omega)\over \sigma_0}={1\over \mu_a \Big(f_0(\infty)\Big)^2}\int^\infty_{U_T}
dU' \, {\alpha\over\sqrt{(U')^5+\alpha^2}}\Big(f_0(U')\Big)^2 e^{-2i\omega\int^{U'}_\infty dU''\,{B(U'')\over C(U'')}}\quad,
\ee
solely in terms of $f_0(U)$.
 \vfil

\end{document}